\documentclass[pre,preprint,epsf,epsfig]{revtex4}
\usepackage{graphicx}
\usepackage{dcolumn}
\usepackage{amssymb}
\usepackage{graphicx}
\usepackage{bm}
\usepackage{epsfig}
\usepackage{float}
\usepackage{psfrag}
\usepackage[utf8x]{inputenc}
\usepackage{ucs}
\usepackage{amsmath}
\usepackage{subfig}
\usepackage{amsfonts}
\usepackage{amssymb}
\usepackage{dcolumn}
\usepackage{latexsym}
\usepackage{hyperref}
\usepackage{caption}

\def\3dots{\:\raisebox{-0.5ex}{$\stackrel{\textstyle.}{:}$}\:}
\def\beq{\begin{equation}}
\def\eeq{\end{equation}}
\def\bea{\begin{eqnarray}}
\def\eea{\end{eqnarray}}
\makeatletter
     \@addtoreset{figure}{section}
\makeatother
\begin{document}

\title{Flocking at a distance in active granular matter}

\author{Nitin Kumar$^1$, Harsh Soni$^{1,2}$, Sriram Ramaswamy$^{2,*}$ and A.K.
Sood$^1$}
\affiliation{$^1$Department of Physics, Indian Institute of Science, Bangalore
560 012, India}
\affiliation{$^2$TIFR Centre for Interdisciplinary Sciences, Tata
Institute of Fundamental Research, 21 Brundavan Colony, Osman Sagar
Road, Narsingi, Hyderabad 500 075, India}
\altaffiliation{On leave from the Department of Physics, Indian Institute
of Science, Bangalore}
\email{sriram@tifrh.res.in}
\date{\today}

\begin{abstract}
The self-organised motion of vast numbers of creatures in a
single
direction is a spectacular example of emergent order. We recreate this
phenomenon using actuated non-living components. We report here that
millimetre-sized tapered rods, rendered motile by contact with an
underlying vibrated surface and interacting through a medium of
spherical beads, undergo a phase transition to a state of spontaneous
alignment of velocities and orientations above a threshold \textit{bead}
area fraction. Guided by a detailed simulation model, we
construct an analytical theory of this flocking transition, with two
ingredients: a moving rod drags beads; neighbouring rods reorient
in the resulting flow like a weathercock in the wind. Theory and
experiment agree on the structure of our phase diagram in the plane of
rod and bead concentrations and power-law spatial correlations near the
phase boundary. Our discovery suggests possible new mechanisms for the
collective transport of particulate or cellular matter.
\end{abstract}
\maketitle

\section*{INTRODUCTION}
\label{intro}
The flocking of animals \cite{vicsekreview,tonertu,mammals,giardina}
relies on organisms sensing each other's presence, orientation and direction of
movement. Physical scientists interested in this spontaneous breaking of
rotation invariance in living systems have adopted a variety of approaches to
the problem \cite{srrmp,srar,srap}, including experiments in vivo and in vitro
\cite{schaller}, computer simulations and theory. Artificial analogues of
self-propulsion \cite{vj,dauchot,kudrolli,sen,golestanian,suropriya,palacci,bricard} are currently a subject of intense interest, both as a controlled
testing ground for theoretical predictions and as a rich source of new phenomena
and applications.

We focus here on the case of macroscopic particles with a
stepwise tapered rod-like shape, with inelasticity and static friction, placed
on a
rapidly vibrating surface. The tilt of the particle takes up the vibrational
energy and, through a frictional mechanism already described elsewhere
\cite{vj,sano,NK,vjthesis}, transduces it into directed motion along the
surface. It is proper to regard such particles as self-propelling: the sustained energy input from the vibrating surface is nutrient or fuel, the tilt of the particle is the motor coordinate, the direction of motion is set by the particle orientation, not by
an external force, and the net in-plane force on particle plus base is zero. The motility mechanism can be viewed as a noisy ratchet \cite{julicherRMP} with the novelty that the asymmetry is carried not by the substrate but by the particle, as in \cite{VKumar}.

A single such motile rod on a bare vibrated surface translates, with mean
speed 0.36 cm s$^{-1}$ and standard deviation 0.25 cm s$^{-1}$, where the
statistics is obtained from 150 rods, primarily in the direction of the vector
from its thick end to its thin end. That vector itself executes rotational
diffusion, losing memory of its initial orientation on long enough time scales.
As a result, the centre of mass of the rod describes a persistent
two-dimensional random walk in the plane of the surface on which it lies.
Anisotropy is crucial: spheres, lacking a distinguished axis, cannot tilt and
thus cannot achieve \textit{motility}, i.e., \textit{self}-propulsion, by means
of the mechanism that animates the rods. A sphere can move laterally only if
pushed or
dragged by the motile rods; an isolated sphere on a vibrating surface simply
bounces in place \cite{NK,vjthesis}. The movements of a single motile rod
amidst a sea of spherical beads were shown earlier \cite{NK} to display some
remarkable statistical properties as a consequence of the obstacles to rod
motion presented by the beads. In the present work we look for novel interaction
effects, structural or dynamical in origin, in the collective behaviour of
motile rods dispersed amongst spherical beads.

Our curiosity is amply rewarded:
we find that the rods
self-organise into a globally oriented moving state at remarkably low
concentration provided the bead background is dense enough. A well-defined
phase transition separates this ordered state or \textit{flock}, from the
disordered isotropic phase. We create a detailed mechanistic simulation model of
this system, which reproduces the phenomena observed and uncovers the
underlying physical processes. We integrate the insights from experiment and
simulation into a simple analytical theory whose predictions are borne out by
our observations. Our work demonstrates for the first time the formation of a
true flock in a collection of dry grains. Our discovery that a small
concentration of motile particles can coherently transport a large passive
cargo, with an efficiency that increases with the load, could find applications
in granular flow. We speculate further that the mutual interaction
of a small population of motile living organisms and a background of non-motile
material could yield self-organisation of the former and enhanced
transport of the latter.

\section*{RESULTS}
\label{results}
\subsection*{Experimental results}
\label{expresults}
\begin{figure}[H]
\centerline{\includegraphics[width=0.6\textwidth]{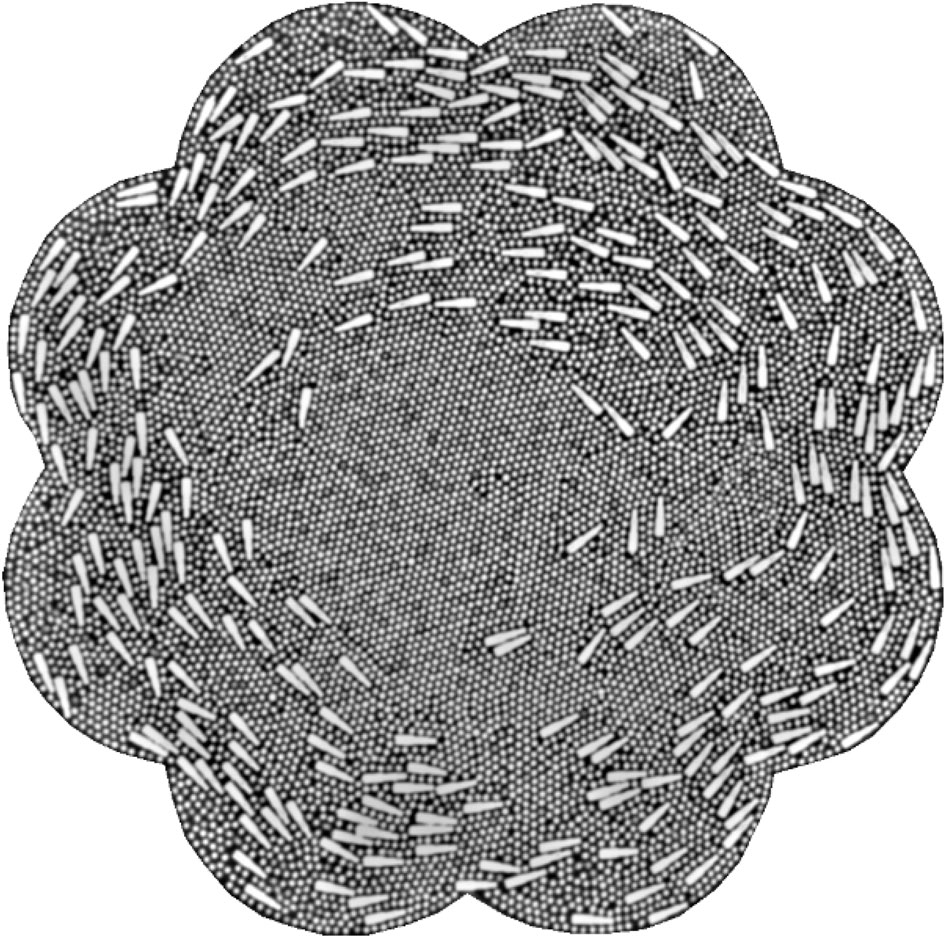}}
\caption{\textbf{A granular flock.} A monolayer of millimetre-sized tapered
brass rods align spontaneously on a vibrating
surface amidst a background of aluminium beads. Bead and rod area fractions
are 0.51 and 0.16 respectively}
\label{flock0}
\end{figure}

\begin{figure}[H]
\centerline{ \includegraphics[width=1.0\textwidth]{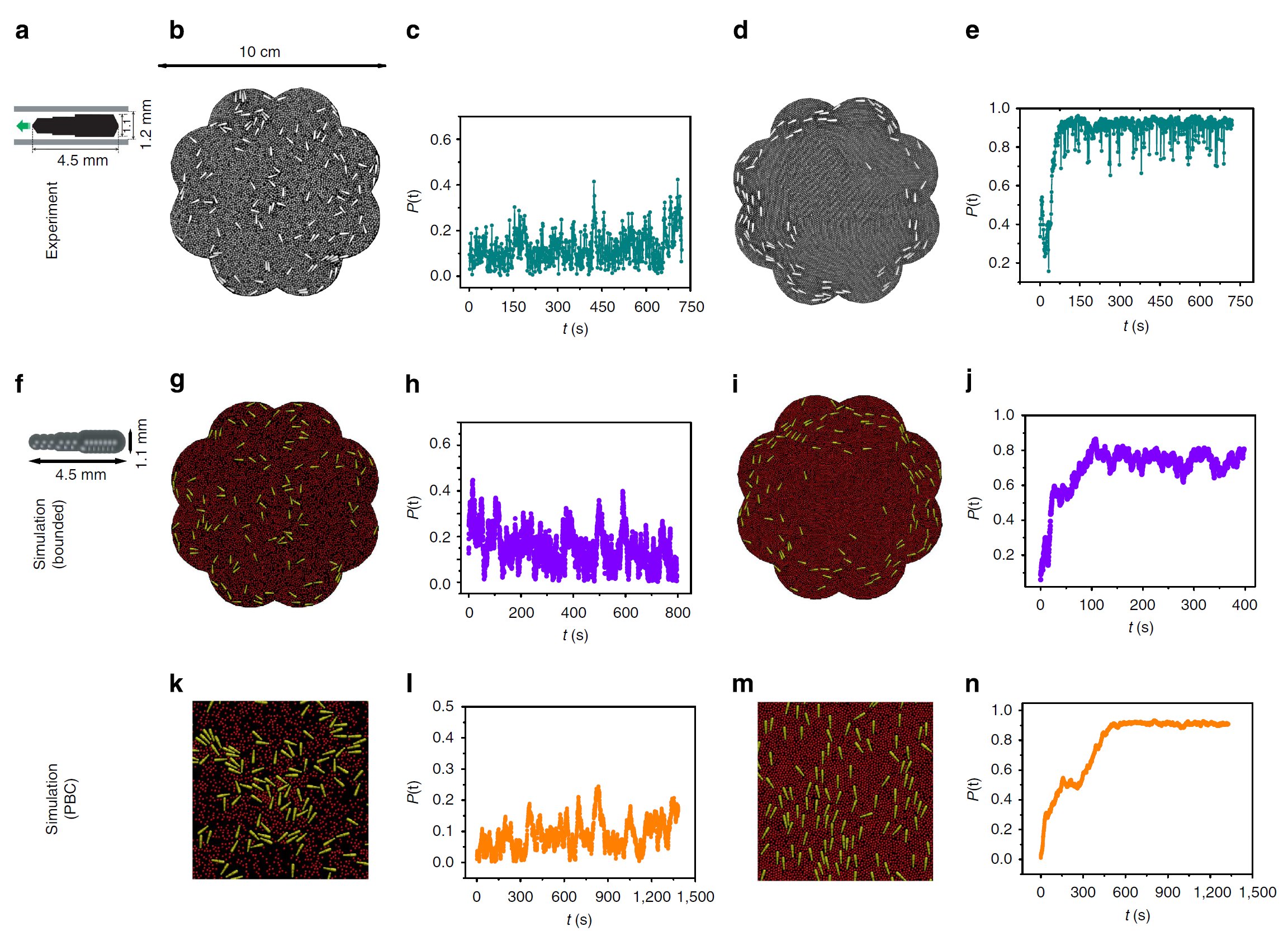}}
\caption{\textbf{Structure and kinetics of the disordered and ordered
states.} (\textbf{a}) Dimensions and confinement geometry of the polar
rod used in the experiment, with arrow indicating direction of self-propulsion.
(\textbf{b}-\textbf{e}): Typical configuration and corresponding time-evolution of polar
order parameter in the disordered phase, bead and rod area fractions $\Phi_b =
0.41$, $\Phi_r= 0.05$ [\textbf{b} and \textbf{c}], and in the ordered phase, $\Phi_b =
0.68$, $\Phi_r= 0.05$ [\textbf{d} and \textbf{e}]; for the simulation in the
flower geometry: \textbf{f} rod used in simulation, (\textbf{g}-\textbf{j}) as in (\textbf{b}-\textbf{e}):
$\Phi_b = 0.48$,
$\Phi_r= 0.06$ [\textbf{g} and \textbf{h}], $\Phi_b = 0.66$, $\Phi_r= 0.06$ [\textbf{i} and
\textbf{j}]; for the simulation with periodic boundary
conditions, configuration and order-parameter evolution for the disordered
phase, $\Phi_b = 0.20$, $\Phi_r= 0.11$ [\textbf{k} and \textbf{l}] and the ordered phase,
$\Phi_b = 0.60$, $\Phi_r= 0.11$ [\textbf{m} and \textbf{n}].}
\label{snapshot}
\end{figure}

Fig. \ref{flock0} shows a representative image of our granular flock. Despite their
low concentration the rod display a high degree of orientational order.
This and further results emerge from our experiments on a mixture of stepwise
tapered rods and spherical beads confined between surfaces subjected to rapid
vertical vibration (see Fig. \ref{snapshot}a and Methods). The system is
initialised by introducing the beads into the sample cell and distributing the
rods amidst the beads, to yield a homogeneous rod-bead
monolayer. We then study the nature of the statistically stationary state of the
vibrated rod-bead mixture as a function of the area fractions $\Phi_b$
and $\Phi_r$, defined as the fraction of the surface covered by the
two-dimensional projections of the beads and rods respectively. We monitor
the centroid positions ${\bf r}_i$ of all the particles and the tail-to-head
unit vectors ${\bf n}_i$, where $i$ labels the particle in question, and all
vectors lie in the horizontal plane.

With increasing $\Phi_b$ (or, less surprisingly, $\Phi_r$) the system is
seen to undergo a phase transition from a disordered state to one in which the
rod orientation vectors are aligned. Correspondingly, the rods move in random
directions in the low-density state (Fig. \ref{snapshot}b and Supplementary
Movie 1) and coherently at high densities (Fig. \ref{snapshot}d and
Supplementary Movie 2). At long times the global movement in the ordered phase
is a clockwise or anticlockwise circulation with equal probability 
(Supplementary Fig. 1), with the orientation vectors aligned
parallel to the boundary. The direction of circulation once selected does not
change for the duration of the experiment. 

We have checked that neither crystalline order nor mechanical rigidity of the
bead bed is an essential ingredient in promoting the flocking
of the rods. For example, the flock in
Fig. \ref{flock0} is on a liquid-like bead
background. Further, rods in an amorphous bead bed, generated by using a
mixture of bead sizes, are also seen to self-organise into a flock
(Supplementary Fig. 2 and Supplementary Note 1). Creating and maintaining a homogeneous 
state in bidisperse systems is complicated by size-segregation effects, which is
why we have continued to work with beads of a single size. The rich
physics that enters when the bead medium is a single- or multidomain crystal
requires a separate study, now underway.

In a confined geometry the simplest flock consists of particles globally
circulating around a common centre, which is the motion seen in our experiments.
Accordingly, we resolve ${\bf n}_i$ into local
polar-coordinate components $({\bf n}_i \cdot \hat{\bf r}_i, {\bf n}_i \times
\hat{\bf r}_i) \equiv {\bf p}_i$, which we use to evaluate the order parameter
${\bf P} \equiv \langle{\bf p}_i \rangle$ distinguishing isotropic and
oriented states, and $G(r) = \langle {\bf p}_i \cdot {\bf p}_j
\rangle_r - P^2$, which
measures correlations of orientational fluctuations about the mean for
pairs $i, j$ of rods with
separation $|{\bf r}_i - {\bf r}_j| = r$.  The
instantaneous in-plane bead-velocity unit vectors ${\bf u}_i$, transformed
to the local polar frame to give $ {\bf v}_i \equiv ({\bf u}_i \cdot \hat{\bf
r}_i, {\bf u}_i \times \hat{\bf r}_i)$, provide another measure of flocking,
${\bf v} \equiv \langle{\bf v}_i \rangle$. The averaging denoted by the angle
brackets
is carried out over all rods and over all times in the steady state for
${\bf P}$, about 100 statistically
independent images for $G(r)$, and 25 frames for ${\bf v}$ (frame rate $=
40$ s$^{-1}$, and the plate is vibrated at 200 Hz).

The disordered and ordered phases are most clearly distinguished by growth
kinetics. Fig. \ref{snapshot}c shows a typical disordered state
with magnitude of the instantaneous order parameter, averaged over rods
but not over time $t$ $P(t) \equiv |{\bf P(t)}|$ fluctuating just above zero,
and Fig. \ref{snapshot}e
presents an
ordered state where $P(t)$ grows and
saturates to a value close to $1$ which
implies that polar rods are now moving parallel to the boundary. We
construct a phase diagram based on the steady-state value of $P$ as
shown in Fig. \ref{PDgr}a. For $\Phi_r > 0.025$ we find that $P$ increases
abruptly across the line
$\Phi_b + k\Phi_r \simeq 0.65$, $k\simeq 2$. We identify this line,
provisionally, as the location of the nonequilibrium phase transition between
the disordered and ordered states, bearing in mind the effects of finite size
and the sample boundary. Correspondingly, Fig. \ref{PDgr}c shows the
experimentally measured polar order parameter as a function of $\Phi_b$ for
$\Phi_r=0.08$. At the
smallest values of $\Phi_r$ the phase boundary bends sharply upwards, possibly
implying a true lower limit to the $\Phi_r$
required to produce a flock regardless of $\Phi_b$.  At the highest densities
arrested states are observed, where rods either jam against the particles of the
bead medium or cluster along the boundary (Supplementary Fig. 3a).
Fig. \ref{PDgr}e reports the variation
of the steady-state orientational
correlations $G(r)$ (scaled by its value at the smallest $r$) of the rods as the
system approaches and then crosses the
phase boundary. For $\Phi_b=0.45$ spatial correlations consistent with a
power-law are seen, while for larger or smaller $\Phi_b$, safely within the
ordered or the disordered phase, correlations decay more rapidly.
These observations can consistently be interpreted as pretransitional
fluctuations in the vicinity of a nonequilibrium phase transition, although the
system sizes explored are too small to permit a claim about the order of the
transition.
As effects of finite size limit the accuracy of our estimates of
the value of the order parameter and the location of the phase
transition, we present in Supplementary Fig. 4a the
correlation functions of the orientation without the
mean subtracted. The flocking of the polar rods is accompanied by coherent
motion of the beads
(Fig. \ref{MediumOP}a)
as well, as measured by the bead-velocity order parameter ${\bf v}$, which
shows the same growth kinetics as $P$ (Fig.\ref{MediumOP}b).

\subsection*{Simulation results}
\label{simresults}
In order to understand the physics underlying this novel realisation of a
flocking transition, and to explore which of the many possible control
parameters are essential, we create a computer model of tapered rods
and spherical beads. For convenience in simulation the rods are constructed
by joining overlapping spheres in a straight line [Fig. \ref{snapshot}f],
but their shape is chosen to resemble as closely as possible the stepwise
tapered rods in the experiment. We have also tried other shapes such as a
uniform taper and find qualitatively similar results which we do not present
here. We carry out two types of
studies, one in an enclosure constructed to imitate the experimental geometry,
the other with periodic boundary conditions in the horizontal plane. In both
cases the simulations, like the experiments, are three-dimensional, with a base
and a lid permitting tightly confined motion in the third dimension. The
energy input, precisely as in the experiment, is through vertical agitation of
the container. The particles move according to the laws of Newtonian mechanics.
Collisions of the particles with each other and with the bounding surfaces are
governed by inelasticity and static Coulomb friction. We add a small rotational
noise to the dynamics of individual rods. This improves the fidelity of our
imitation of the experimental system, in which minute imperfections in particle
shape and base or lid topography endow each rod with rotational diffusion even
when no beads or other rods are present. 

\begin{figure}[H]
\centerline{\includegraphics[width=0.7\textwidth]{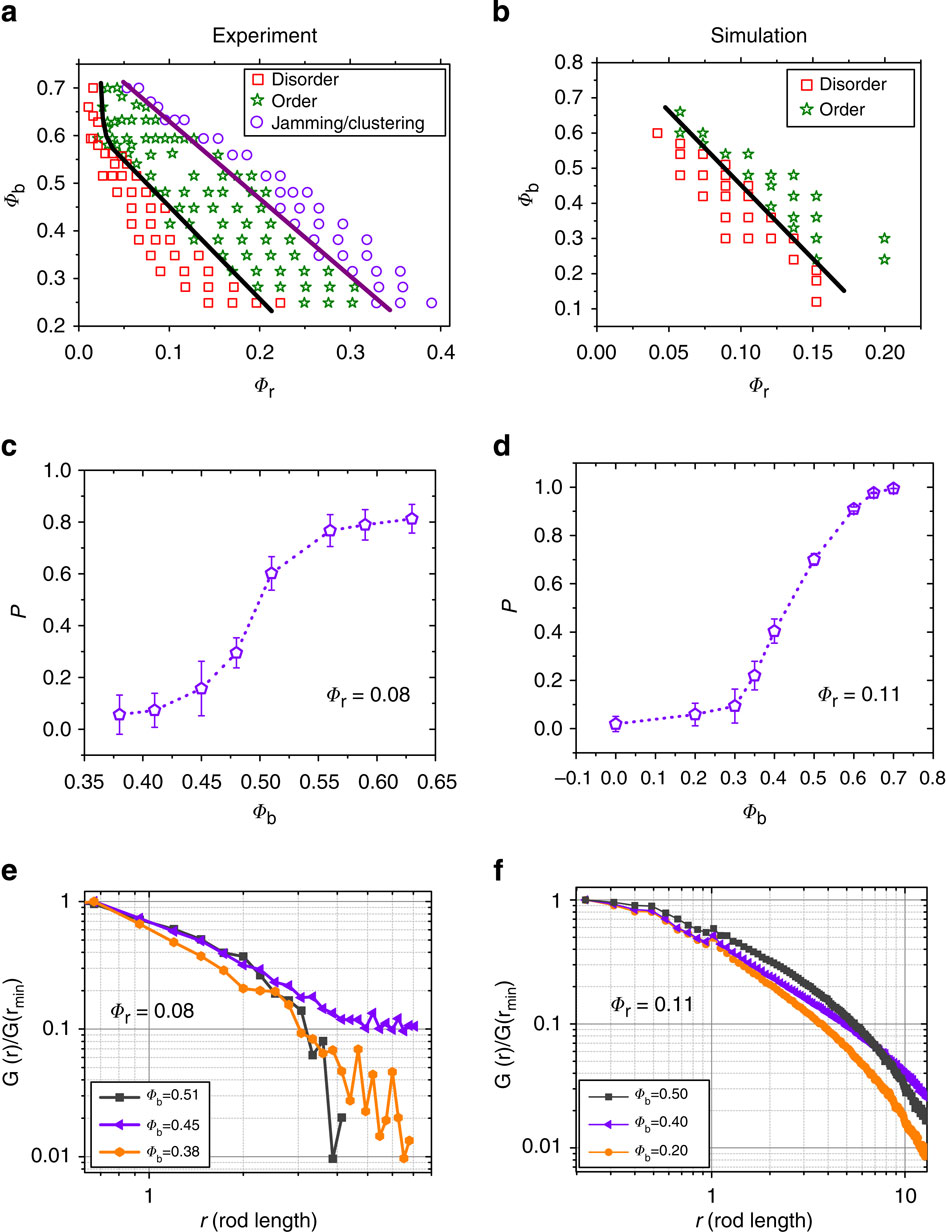}}
\caption{\textbf{Flocking phase transition and growth of correlations.}
(\textbf{a}) Experimental phase diagram in the $\Phi_b$-$\Phi_r$
plane, showing the phase boundary between the isotropic phase and the flock.
Clustering at the sample boundary sets in at high concentration. (\textbf{b})
Simulation phase diagram in the flower geometry reproduces the flocking
transition. We have not explored concentrations at which boundary clustering is
expected. Increase in polar order parameter $P$ as
function of $\Phi_b$ for experiments (\textbf{c}) and simulations
(\textbf{d}). The error bars correspond to the standard deviation of $P$.  Orientational correlation function from experiments (\textbf{e})
and
simulations in a 55.1-rod-length periodic box (\textbf{f}) at various $\Phi_b$
at a fixed $\Phi_r$. A considerable range of power-law correlations is
seen as $\Phi_b$ increases.}
\label{PDgr}
\end{figure}

\begin{figure}[H]
\centerline{\includegraphics[width=0.9\textwidth]{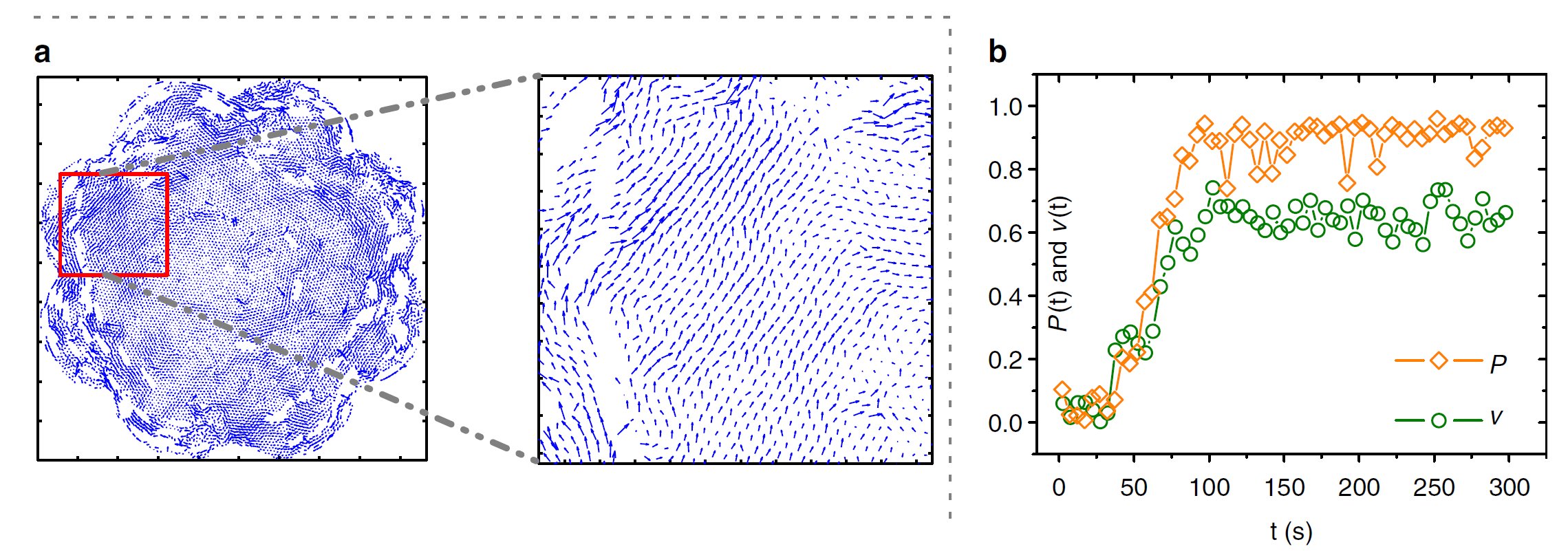}}
\caption{ \textbf{Correlation between the dynamics of the polar rods and the
bead medium.} (\textbf{a}) Arrows in the zoom show the coherent velocity field
of the bead medium in the ordered phase with $\Phi_b=0.66$ and $\Phi_r=0.06$. Blank regions are occupied by polar rods. 
(\textbf{b}) The
ordering kinetics of the bead velocity field
tracks
that of the polar orientation.}
\label{MediumOP}
\end{figure}

Our simulation results can be seen in Figs. \ref{snapshot}g-j \& k-n and
Fig. \ref{PDgr}. In the ``flower'' geometry
we find precisely the behaviour seen in the experiment. Random motion at low
area fraction (Supplementary Movie 3) gives way to an ordered, circulating flock at high area fraction (Supplementary Movie 4),
and a phase diagram broadly corresponding to that in the experiment is obtained.
The
numerical experiment, moreover, can be performed in a periodic simulation box,
eliminating the effects of sample boundaries. Fig. \ref{snapshot}k \& l
(Supplementary Movie 5) and \ref{snapshot}m \& n (Supplementary Movie
6) show
that the order parameter, now measured in a global Cartesian frame, remains at
values consistent with zero for all times at low $\Phi_b$, while at high
$\Phi_b$ it grows to saturation. Clear evidence of a flocking transition
around $\Phi_b \simeq 0.35$ is seen in Fig. \ref{PDgr}d showing polar order
parameter as a function of bead area fraction. Consistent with these findings
and with the experiments, Fig. \ref{PDgr}f shows evidence of power-law
correlations for $\Phi_b=0.4$, but appreciable curvature on a log-log plot for
$\Phi_b$ well in the ordered or the disordered regime. Increasing
$\Phi_b$ serves to promote orientational correlations for $\Phi_r$ as low as
0.03 (Supplementary Fig. 5). As in the experiments, we
present the ``unsubtracted'' correlation functions in Supplementary Fig. 4b. For good measure, we
also investigate and rule out, in Supplementary Fig. 6 (and Supplementary Note 2), the possibility
that the only role played by the beads is to reduce the rotational diffusion of
rods. These simulation results establish beyond reasonable doubt that the
phenomenon we observe is a phase transition, with the beads playing a direct
role in mediating an aligning interaction amongst the rods.

\begin{figure}[H]
\centerline{\includegraphics[width=1.0\textwidth]{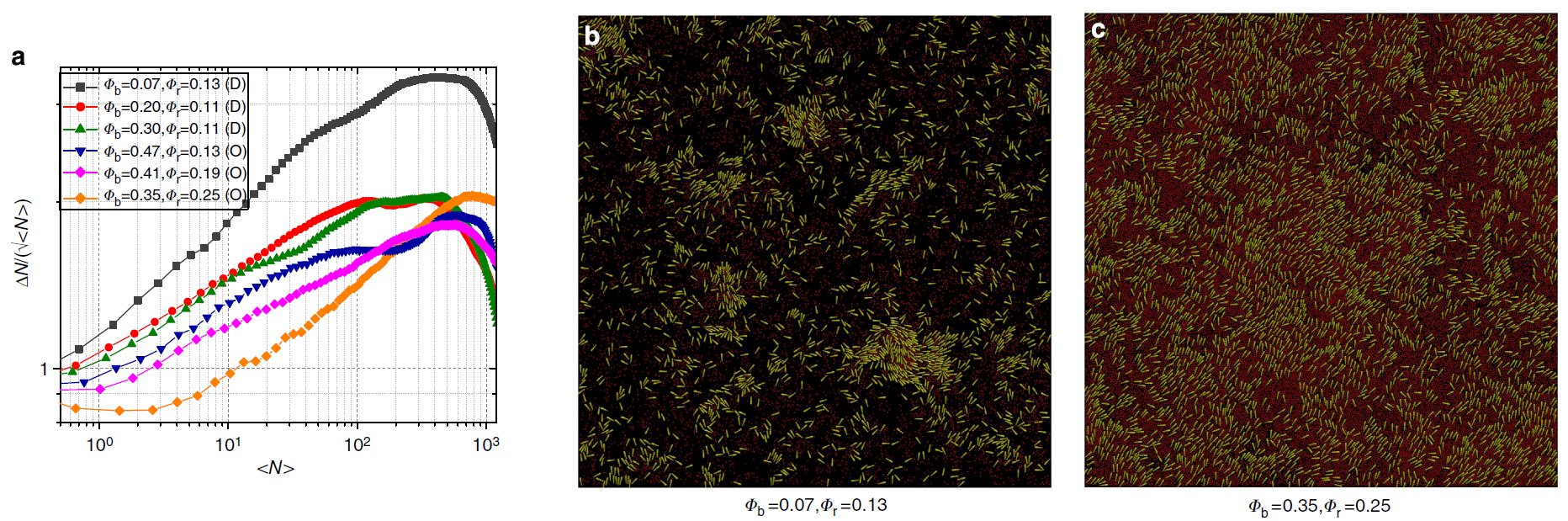}}
\caption{ \textbf{Enhanced density fluctuations.} (\textbf{a}) The scaled
standard deviation $\Delta N/\sqrt{\langle N \rangle}$
in the number of polar rods, for regions containing $\langle N
\rangle$ rods on average, is seen to grown with $\langle N \rangle$ for area
fractions in the ordered (labelled O) and disordered (D) phases. (\textbf{b})
and (\textbf{c}):
Representative images from the two phases.}
\label{denfluc}
\end{figure}

Lastly, on theoretical grounds, in particular because the suspending medium of
beads is compressible \cite{bricard}, we expect large number fluctuations in our
system as in standard flocking models of the Toner-Tu type \cite{tonertu}. In
such systems, large-scale correlated particle currents arise as a result of the
coupling to easily excited and slowly decaying variations in the orientation
field, both in the ordered phase \cite{tonertu} and in the isotropic phase near
onset \cite{fily}. In Methods we show that the separate conservation of rods and
beads leaves unaltered the arguments \cite{tonertu} for anomalously large
number fluctuations. We have carried out a limited set of numerical studies of
fluctuations in the number density of rods in the ordered and disordered phases,
and find an excess in both cases (Fig. \ref{denfluc}).  Our simulation
measurements of the standard deviation $\Delta N$ in the number of polar rods in
regions containing $\langle N \rangle$ particles on average show that $\Delta
N/\sqrt{\langle N \rangle}$ grows with $\langle N \rangle$. However,
more extensive simulations or experiments are required for a useful
fit to a power law or other dependence on $\langle N \rangle$, and to probe the
relation of this phenomenon to known sources
\cite{tonertu,srrmp,fily,cates,frenkel} of enhanced density fluctuations in
active systems.

\subsection*{Theory}
\label{theory}
\begin{figure}[H]
\centerline{\includegraphics[width=1.0\textwidth]{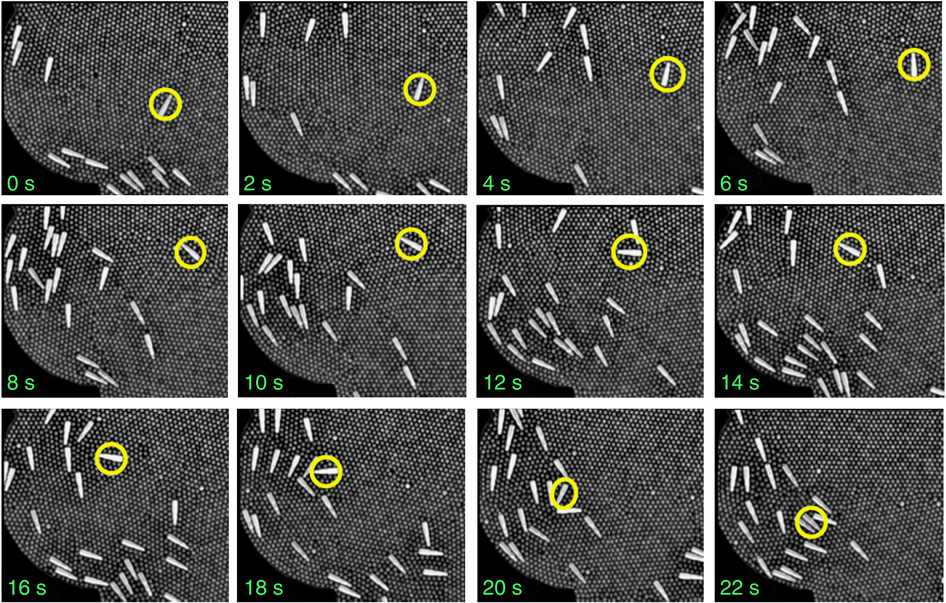}}
\caption{\textbf{Flowing beads reorient polar rods.} A series of experimental
images showing a stray polar rod brought
into alignment with the rest of the flock as a consequence of bead flow.}
\label{notouch}
\end{figure}

We now integrate the insights gained from experiment and simulation to construct
a theory. The sequence of events in Fig. \ref{notouch} also seen in simulation
studies, shows that coherently moving polar rods are able to entrain a stray rod
initially moving in the opposite direction, through the bead flow they generate, as
seen in Supplementary Movies 2, 4 and 6. These observations suggest that the
coupling of flow and orientation lies at the heart of the phenomenon of flocking
at a distance, as we now show through a minimal, universal hydrodynamic theory 
whose parameters and their dependence on $\Phi_b$ we measure in our
simulation. We also discuss the
connection to the particular case of Bricard \textit{et al.} \cite{bricard}. 

As we are concerned with large-scale ordering, we work with coarse-grained
fields. Our system
is a suspension of polar rods in a \textit{compressible} two-dimensional fluid
of beads on a substrate. The variables of interest are therefore the local order
parameter ${\bf P}({\bf r},t)$, given by the average of the orientation vectors
${\bf p}_i$ of the polar rods, and the number densities $\rho$ and $\sigma$ of
beads and rods respectively, in a small neighbourhood around point ${\bf r}$ at
time $t$. Below, for simplicity, we consider the limit of small rod
concentration, and so treat only the bead density $\rho({\bf r},t)$. A
treatment with both densities is in Methods. In addition,
we include ${\bf v}({\bf r},t)$, the average bead-velocity vector in the
neighbourhood, as it plays a crucial role in mediating the interactions that
lead to order. In the dilute-rod limit ${\bf v}$ can be viewed as the total
velocity field of both species. 

The equations of motion follow from a few general principles and some key
elements of the dynamics of the system. Conservation of particles is embodied in
the continuity equation
\beq
\label{conteq}
\partial_t \rho + \nabla \cdot (\rho {\bf v}) = 0. 
\eeq
Newton's second law locally reads
\beq
\label{momeq}
\rho \partial_t {\bf v} = -(\Gamma - \eta \nabla^2){\bf v} + \alpha {\bf
P} - B \nabla \rho + ...
\eeq
where we present only the leading-order terms in an expansion in powers of
gradients and fields. In \eqref{momeq}, $\Gamma>0$ damps motion with respect
to the substrate and lid, the term in $B$ can be viewed as describing pressure
forces, driving flow downhill in density for $B>0$, and $\alpha$ measures the
degree to which orientation of the polar rods gives rise to propulsive
flow. We note in passing here that such a term was argued to arise
\cite{aranson} in the two-dimensional projection of the three-dimensional
hydrodynamics of a suspension of motile organisms confined to a thin layer of
fluid. We discuss below the somewhat different origin of the term in the case of
our monolayer system. The viscosity $\eta$ describes transmission of
momentum in the plane of the system. The local order
parameter ${\bf P}$, again to leading
order in gradients and fields, obeys
\beq
\label{Peq}
\partial_t {\bf P} = \lambda{\bf v} - (a - K \nabla^2){\bf P} - A \nabla
\rho + ...
\eeq
where $a$ measures the rate at which an initial polarisation relaxes,
presumably through rotational diffusion, and the magnitude and sign of $A$
characterise the tendency of rods to orient parallel or antiparallel to
concentration gradients \cite{tonertu,saunders}. The parameter
$\lambda$
determines the rate at which a uniform velocity (with respect to the confining
walls) aligns the polarisation. The parameter $K$ governs spatial
variations in the direction and magnitude of the order parameter. The ellipsis
in \eqref{momeq} and \eqref{Peq}
denotes contributions \cite{tonertu} at higher orders in ${\bf p}$, $ {\bf v}$
and $\nabla$,
discussed in Methods.

The terms with coefficients $\alpha$ and $\lambda$ are the key players in
\eqref{momeq} and \eqref{Peq}. No symmetry rules out their existence; a
phenomenological theory must therefore include them. However, the physics
underlying these terms, especially as regards their \textit{signs}, merits some
discussion. Recall that ${\bf p}_i$ for the $i$th rod points from the
thick to the thin end. Depending on the detailed contact mechanics with the
vibrating base, the rod could propel itself on average parallel or antiparallel
to ${\bf p}_i$. Friction would cause the bead medium in the vicinity to be
dragged in the direction of motion of the rod. A collection of such rods thus
generates a force proportional to the mean polarisation ${\bf P}$. The
parameter $\alpha$ measures the strength of this forcing. The
magnitude of $\alpha$ should grow with the mean rod concentration and propulsion
speed, and should further depend on the surface characteristics of a rod which
influence its ability to carry the ambient medium with it, by dragging or
pushing. A rod with spontaneous motion parallel
(antiparallel)
to ${\bf p}_i$ has $\alpha > 0$ ($\alpha < 0$). By suitably engineering the
geometry, mass distribution, surface properties and contact mechanics of a rod,
and exploring a range of vibration parameters, it should be possible to make
polar rods with a range of magnitudes and either sign of $\alpha$; the rods we
work with here have $\alpha > 0$.

\begin{figure}[H]
\centerline{\includegraphics[width=.6\textwidth]{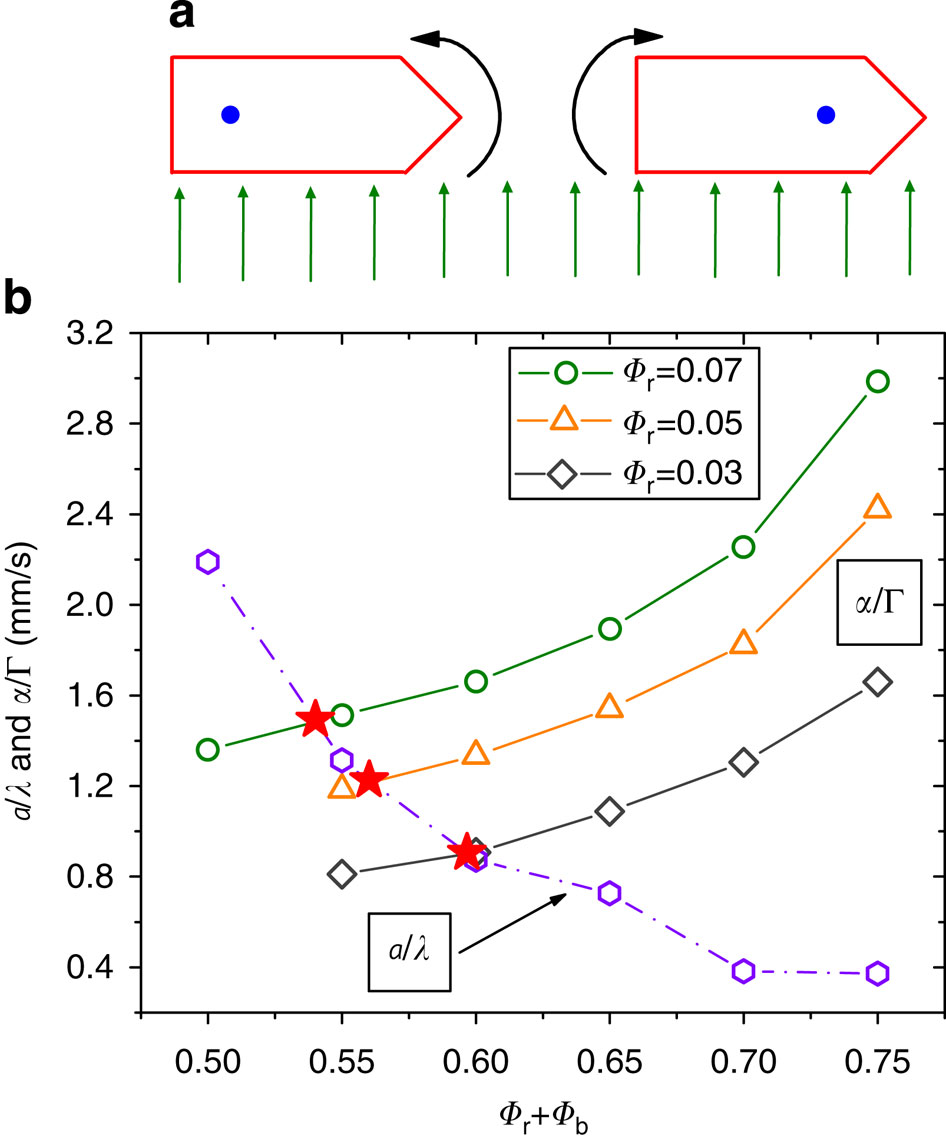}}
\caption{\textbf{Weathercock effect and location of the phase transition.}
(\textbf{a})
The direction in which the polar rod is rotated by a flow depends on its natural
pivot point. (\textbf{b}) $\alpha/\Gamma$ is seen to grow and $a/ \lambda$
to decrease with increasing
$\Phi_b$. Thus the flow velocity generated by a local polarisation as well as
the aligning power of bead flow grow with increasing bead concentration.
The mean-field estimate of the flocking transition corresponds to the
intersection of the two curves, marked by stars, where $\bar{a}$ in Eq. \eqref{Peqeff}
changes sign from positive to negative as $\Phi_r$ + $\Phi_b$ is increased.}
\label{parameters}
\end{figure}

The mechanics underlying $\lambda$ is reminiscent of the interaction of wind
with a weathercock. Imagine a single tapered rod lying athwart a uniform
flow of particles over a surface. Although the momentum transfer is distributed
uniformly across the rod, its mass is concentrated, and hence its pivot point
shifted, towards one end. This leads to a net torque due to the flow, and a
consequent rotation as shown in Fig. \ref{parameters}a. Geometrical arguments
alone cannot determine the sign of $\lambda$ and hence the sense of rotation,
and self-propelling activity could influence its value. It should be possible to
engineer $\lambda < 0$ by using intrinsically higher-density material at the
thin end, or a hollow interior at the thick end.

The availability of a numerical simulation model provides independent
measurements of all phenomenological parameters of
importance to our theory. To measure $\lambda$, we place a
single polar rod
initially pointing in the $x$ direction, and impose a flow of beads with
velocity ${\bf v}$ in the $y$ direction (See Supplementary Movie 7). By measuring the initial growth rate
we estimate the parameter $\lambda$ in \eqref{Peq}, which we average over 50
such trials. Independently, by measuring the rotational relaxation of a single
rod, we infer $a$. In a separate simulation, we impose a nonzero mean ${\bf P}$
by applying an orienting field on a collection of rods. Measuring the saturation
value of the resulting macroscopic velocity yields $\alpha/\Gamma$ in 
\eqref{momeq}. The damping $\Gamma$ is readily estimated from the relaxation of
an imposed initial velocity. Fig. \ref{parameters}b shows the dependence of these
key parameters on area fraction $\rho_0$. We comment in Methods on the
parameter $A$, which plays only a minor role in our analysis.

We ignore density variations in (\ref{Peq})
and (\ref{momeq}), replacing $\rho$ by its mean value $\rho_0$ which we
treat as a parameter on which the coefficients depend. Working with
the spatial Fourier components ${\bf v}_{\bf q}$ and ${\bf P}_{\bf q}$ of the
velocity and order parameter at wavevector ${\bf q}$, (\ref{momeq}) then tells
us that, on time scales long compared to $\rho_0/\Gamma$,
${\bf v}_{\bf q} \to \alpha {\bf P}_{\bf q} / (\Gamma + \eta q^2)$.
Substituting this value for ${\bf v}$ in
(\ref{Peq}) we obtain, to order $q^2$, 
\beq
\label{Peqeff}
\partial_t {\bf P} = - (\bar{a} + \bar{K}q^2){\bf P}
\eeq
where $\bar{a} = a - \lambda \alpha / \Gamma$  and $\bar{K} = K + \lambda
\alpha \eta /\Gamma^2$. The mechanism for spontaneous
ordering now becomes clear. As $\lambda$ is an increasing function of $\rho_0$,
$\bar{a}$ can turn negative even if $a$ is positive. The state ${\bf P} = 0$ is
then linearly unstable, and a state with nonzero ${\bf P}$ on macroscopic scales
will set in. Let us now test our theory using our simulation measurements
of the parameters. First, for the proposed mechanism to work, $\bar{a}$ must
be less than $a$, i.e., we predict $\lambda \alpha / \Gamma > 0$. Our
simulations show that a rod drags beads in the direction of its own motion and
that the flow of beads rotates neighbouring rods to point in, and thus move
in, the direction of the flow. Thus $\alpha$ and $\lambda$ have the same sign,
confirming the prediction. Secondly, Fig. \ref{parameters}b shows that as
$\Phi_r + \Phi_b$ is
increased, $\alpha/\Gamma$ (for three values of $\Phi_r$) increases
and $a/\lambda$ decreases, and the two quantities cross (i.e., $\bar{a}$
crosses zero) at a value of $\Phi_r + \Phi_b$ which decreases with increasing
$\Phi_r$. Therefore the theory predicts a phase transition to an ordered phase
with increasing area fraction, and negative slope to the phase boundary in the
$\Phi_r$-$\Phi_b$ plane, both of which are borne out by experiment. The
intersection points in Fig. \ref{parameters}b, corresponding to the mean-field estimate of
the threshold area fraction, are, not surprisingly, somewhat lower than the
measured values in the simulation, Fig. \ref{PDgr}b. 

Within our mean-field treatment [(\ref{Peqeff}) or the parent
equations (\ref{Peq}) and (\ref{momeq})] it is
straightforward to see that the transition has a continuous onset and
a diverging correlation length $\sim \sqrt{\bar{K}/\bar{a}}$. Our
limited observations in experiment and simulation show an increasing
correlation length upon approach to the transition. Experience with standard
flocking models would suggest \cite{chaterefs} instead a discontinuous
transition, albeit with appreciable pretransitional fluctuations. Larger-scale
simulation studies including finite-size scaling analyses, now in
progress, will determine the character and spatial structure of the onset of
order for our rod-bead flock. Note that orientability, motility and coupling to
flow, rather than specifics of shape, were the essential ingredients of our
theory. On symmetry grounds one cannot therefore rule out the possibility that
the self-propelled disks of \cite{dauchot}, in a background of non-motile
particles, could flock through the mechanism reported here -- provided flow can
reorient the disks in the manner required.

\section*{DISCUSSION}
\label{discussion}

Our work is of importance as a remarkable example of cooperativity in
active matter, with rods and beads giving rise to order that the former
could not achieve at such low concentration and the latter could not
aspire to at all. It suggests an original mechanism for creating a
coherent motile composite by introducing a small number of orientable
motile particles into a medium containing non-motile components, that
transmit motion more effectively as their concentration is increased. A
static image of the ordered state would give no clue to why the rods are
lining up: the transmission of orientational information takes place
through a viscous interaction which can be tuned through concentration.
This works because our system is a monolayer, not a confined
three-dimensional fluid. Momentum transfer in the vertical direction is
governed by friction with the bounding base and lid. Only the
transmission of momentum laterally in the plane of the system is
governed by an effective viscosity arising from tangential forces
between the beads with each other and with the rods. Our simulations
show that in-plane viscous effects, as measured by $\alpha$ or $\eta$ in
(\ref{momeq}), are enhanced by increasing bead concentration, while the
out-of-plane damping $\Gamma$ stays constant. It is this key feature that gives
us a system in which the flocking tendency can be tuned by varying the
concentration of the non-motile actors, the beads.

Empirically, the system displays a flocking transition even if the bead
medium is so concentrated that it crystallises. Our simplified theory is
restricted to the case where the system as a whole is fluid, not
crystalline. How motile rods move through, and transmit their motion and
orientation through, a highly ordered bead medium, remains a major open
problem.

The hydrodynamic theory we have presented is indifferent to the precise
origin of motility. It matters only that a polar, motile object drags
ambient inert particles, whose flow in turn reorients neighbouring polar
particles. There is a priori no reason why this cannot be realised in a
system in which the motility mechanism does not required mechanical
vibration or, for that matter, a vertical electric field as in
\cite{bricard}. Could living matter take advantage of such a general
mechanism?

The reader should note that the flocking behaviour seen in our system
was not inevitable, but relied on favourable signs of the parameters
$\alpha$ (rods drag beads) and $\lambda$ (the weathercock effect). An
important direction that we are currently exploring is to design polar
rods with the ``wrong'' relative sign of $\alpha$ and $\lambda$. Indeed
all the parameters $\alpha, \, \Gamma, \, \lambda$ and $a$ can be
manipulated at the scale of a single particle. Increasing rotational
noise, perhaps by roughening particle surfaces or increasing the
vibration amplitude, should increase $a$. $\Gamma$ could be changed by
roughening the base and lid. The tendency of a polar rod to point along
the flow is quantified by $\lambda$. The shape and the mass distribution
of the rod are the likely properties governing this parameter. It seems
to us that there are two mechanisms behind the term $\alpha \textbf{P}$
in \eqref{momeq}: dragging by tangential forces at the sides of a rod,
and the pushing of beads at the head of a rod. A detailed exploration of
the effects of shape and surface friction would allows us to manipulate
$\alpha$.

Much remains to be explored about this granular flock. Deep in the
ordered phase, the character of excitations about the uniform aligned
state awaits study, as do the statistics of density and orientational
fluctuations. Just past the apparent onset of order, does our system
reproduce the banding instability of simple flocks, and is the character of
this granular flocking phase transition discontinuous \cite{chaterefs}? What is
the
nature of the competition between kinetic arrest and flocking as concentrations
are increased?

Before closing, we return to the colloidal flock of Bricard \textit{et
al.}
\cite{bricard}, in which energy from a vertical electric field is
transduced into rolling motion in a horizontal plane. In that system
too, hydrodynamic coupling leads to a macroscopically aligned state of
particle velocities. The most important distinguishing feature of our
system is that the aligning interaction between the motile polar rods
can be tuned by changing the concentration of the beads, which are
themselves non-motile. Furthermore, the suspending medium, namely, the
fluid of beads, is compressible (effects associated
with overall incompressibility of grains plus air should arise in an
airtight container.), which rules out the suppression of
fluctuations found by Bricard \textit{et al.} \cite{bricard}; the rods
carry an intrinsic shape polarity; the single-rod motility is not
hydrodynamic in origin; the hydrodynamics at all scales in our monolayer
system is that of a two-dimensional fluid on a dissipative substrate
\cite{srmaz}; the flow field around a single polar rod (Supplementary Fig. 7 and Supplementary Note 3) is simply a  screened monopole \cite{brotto}.

To summarise: we have created a system of macroscopic motile dry grains
that flock
spontaneously at low concentration, transmitting information about their
orientation over large distances through the flow of an ambient medium
of millimetre-sized beads. The beads, themselves non-motile, are cargo
and coupling: increasing their number promotes flocking and enhances
their own coherent transport. The simple and robust mechanism underlying
this nonequilibrium phase transition prompts us to wonder about its possible
wider relevance in living matter and industry.

\section*{METHODS}
\label{methods}
\subsection*{Experimental methods}
\label{expmethods}
Our experiments are conducted on a mixture of macroscopic brass rods and
aluminium
beads. The rods \cite{NK} are 4.5 mm long, with diameter tapered in steps from
1.1 mm at
the thick end or \textit{tail} to 0.7 mm at the thin end or \textit{head}, as
shown in Fig. \ref{snapshot}a. The beads are spheres of diameter 0.8 mm.
The rod-bead mixture is confined to a monolayer in an experimental cell
consisting of a shallow circular well in an aluminium plate
whose dimensions and shape are shown in Fig. \ref{snapshot}b. The plate  is
covered by a glass lid, and the distance $w$ from plate to lid is 1.2 mm. The
assembly is vibrated vertically by a magnetic shaker (LDS V406-PA100E).
All our experiments are carried out
at oscillation amplitude $\mathcal{A}= 0.04$ mm and frequency $f=200$ Hz,
corresponding to
a nondimensional shaking strength $(2 \pi f)^{2} \mathcal{A} /{g}=7.0$ where
$g$ is the acceleration due to gravity.
The rods imitate self-propulsion by
transducing the vertical vibration into fluctuating but persistent motion in the
plane, in the tail-to-head direction \cite{sano,vj,NK,vjthesis} [the arrow
in Fig. \ref{snapshot}a].
We prevent accumulation of particles at the
boundary,
seen in Supplementary Fig. 2b, by the
use of a flower-shaped sample cell \cite{dauchot}.
Images of the collection of particles were acquired by high-speed camera (Basler
acA2040-180 kc, $2046\times2046$ pixels giving a spatial resolution of 0.05 mm
for our setup) at 40 frames per second and analysed in ImageJ (http://rsb.info.nih.gov/ij/).

\subsection*{Simulation methods}
\label{simmethods}
Our simulation is time-driven, as particle shapes and the vibrating base and lid
complicate the prediction of the next collision. All the interactions in the
simulations are based on the Impulse Based Rigid Body Collision Model
\cite{collision1,collision2}. Each tapered rod consists of equally
spaced $13$ overlapping spheres, seven of $1.1$ mm diameter and three each of
$0.88$ mm and $0.72$ mm (Fig \ref{snapshot}f), with mass density 8.7 gm/cm$^3$
corresponding to brass. The beads are represented by spheres of diameter $0.8$
mm and mass density 2.7 gm/cm$^3$ of aluminium. The vertical positions of
the vibrating base and lid at time $t$ are $\mathcal{A} \cos 2 \pi f t
+\mathcal{A}$ and $\mathcal{A} \cos 2
\pi f t + \mathcal{A}+ w$ respectively. All quantitative studies with periodic
boundary
conditions (PBCs) are all for a system size of 55.1 rod lengths; some videos
use a different size. The rods in the experiment display rotational
diffusion as a result of slight imperfections in shape or substrate
roughness. Lacking such asperities, our numerical rods need to be supplied
with rotational noise: if a rod collides with the bed or the lid with relative
velocity $ v_{rel}$ of contact points normal to contact plane, an extra angular
velocity $\omega_{z}=\varepsilon v_{rel} \eta$ is supplied to the rods. Here
$\eta = \pm 1$ with equal probability. The strength $\varepsilon$ is a control
parameter. In order to match closely the experimentally observed single-particle
dynamics, we choose the friction and restitution coefficients $\mu$ and
$e$ to be 0.05 and 0.3 for particle-particle collisions, 0.01 and 0.3 for
bead-wall (base, lid and lateral boundary) collisions, 0.03 and 0.1 for
rod-base(or rod-lid) collisions, 0.01 and 0.3 for rod-boundary collisions
respectively. Between collisions the particles obey Newtonian rigid body
dynamics. All simulation movies and snapshots are generated using VMD
software \cite{VMD}.

\subsection*{Details of theoretical development}
We construct here a hydrodynamic description of the fluid of beads and motile
polar rods on a substrate, keeping track of the separately conserved numbers
of beads and rods. We show that the flocking transition mediated by the bead
velocity is unaltered by the additional conservation law. We demonstrate in
addition that the presence of the dense fluid of beads leaves unaffected the
singular nature \cite{tonertu} of number fluctuations in a flock. This is in
contrast to the case \cite{bricard} of active particles in a confined
incompressible fluid medium.

Let $\rho$ and $\sigma$ be the bead and rod
concentration fields respectively, and let ${\bf v}$ be the bead velocity field.
Let us identify the polar-rod current with the polar order parameter ${\bf p}$
as is frequently done in microscopic derivations of the flocking equations
\cite{bertin}. Extending our arguments in the main body of the paper, the
governing equations of motion, a two-component extension of the Toner-Tu
\cite{tonertu} model,  read
\beq
\label{rodcont}
\partial_t \sigma + \nabla \cdot {\bf p} = 0,
\eeq
\beq
\label{beadcont}
\partial_t \rho + \nabla \cdot \rho {\bf v} = 0,
\eeq
expressing conservation of beads and polar rods respectively,
\beq
\label{beadmom}
\rho \partial_t {\bf v} + \Gamma {\bf v} + O(\nabla \nabla {\bf v}) + ... =
\alpha
{\bf p} - E \nabla \rho - F \nabla \sigma
\eeq
which is Newton's second law for the bead fluid, with damping $\Gamma$ due to
contact with the substrate and lid, propulsion $\alpha$, and moduli
$E$ and $F$ giving rise to pressure-like forces from bead and rod
density gradients, and
\beq
\label{polarfull}
\partial_t {\bf p} + w {\bf p} \cdot \nabla {\bf p} + ... = -(a + b {\bf p}
\cdot {\bf p}){\bf p} + (K_1 \nabla \nabla \cdot + K_3 \nabla^2) {\bf p} +
\lambda {\bf v} - A \nabla \rho - B \nabla \sigma
\eeq
describing the dynamics of the polar order parameter and introducing several
phenomenological coefficients. We have ignored two other terms with two factors
of ${\bf p}$ and a single $\nabla$, which are unimportant to our analysis.
The ellipsis in (\ref{beadmom}) and (\ref{polarfull}) denotes terms such as
$\nabla \nabla {\bf p}$,
higher orders in ${\bf p}$, and \cite{tonertu} order ${\bf p} \nabla {\bf p}$,
as well as terms obtained by substituting one or more factors of ${\bf p}$ in
the preceding by ${\bf v}$. These include conventional viscous damping $\nabla
\nabla {\bf v}$. None of these has any significant effect on our analysis.

On timescales
$\gg \rho/\Gamma$ (\ref{beadmom}) tells us ${\bf v}$ relaxes to a value
determined by the remaining fields. Inserting this value in (\ref{beadcont})
and (\ref{polarfull}) yields effective equations of motion for $\rho$ and ${\bf
p}$, in which (\ref{polarfull}) retains its form but with $a \to \bar{a} = a -
\lambda \alpha / \Gamma$, $A \to \bar{A} = A + \lambda E / \Gamma$ and $B \to
\bar{B} = B - \lambda F / \Gamma$. For large $\lambda \alpha$
$\bar{a}$ turns negative, leading to a flock with mean order parameter
${\bf p}_0$, with magnitude $p_0 = \sqrt{-\bar{a}/b}$ and mean bead velocity
${\bf v} = (\alpha / \Gamma) {\bf p}_0$. Let us denote the direction of mean
orientation by $||$ and perpendicular directions as $\perp$. Expanding to linear
order about this state, ${\bf p} = {\bf p}_0 + \delta {\bf p}_{\perp}, \, {\bf
v} = {\bf v}_0 + \delta {\bf v}_{\perp}$, $\rho = \rho_0 + \delta \rho$,
$\sigma = \sigma_0 + \delta \sigma$
and defining
\beq
\label{Theta}
\Theta = \nabla \cdot \delta {\bf p}_{\perp}
\eeq
yields
\beq
\label{ptrans}
(\partial_t + v_1 \partial_{||} - K \nabla^2) \Theta = - \bar{A}
\nabla_{\perp}^2 \delta \rho  - \bar{B} \nabla_{\perp}^2 \delta \sigma
\eeq
for the orientation, with $v_1 = w p_0$,
\beq
\label{beadord}
(\partial_t + v_0 \partial_{||} - {\rho_0 E \over \Gamma}  \nabla^2) \delta \rho
+{\rho_0 \alpha \over \Gamma} \Theta - {\rho_0 F \over \Gamma} \nabla_{\perp}^2
\delta \sigma = 0
\eeq
for the concentration of beads, and
\beq
\label{rodcontord}
\partial_t \delta \sigma + \nabla \cdot {\delta \bf p} = 0
\eeq
for  the rods. Fourier-transforming in space, and eliminating
$\sigma$ in (\ref{ptrans}) and (\ref{beadord}) via (\ref{rodcontord}) yields
\beq
\label{rhodiff}
\partial_t^2{\delta \rho}_{\bf q} - (-iv_0 q_{||} + {\rho_0 E \over
\Gamma}q^2)\partial_t \delta \rho_{\bf q} -  {\rho_0 \alpha \over \Gamma}
\partial_t {\Theta}_{\bf q} + {\rho_0 F \over \Gamma}q^2  \Theta_{\bf q}
\eeq
and
\beq
\label{Thetadiff}
\partial_t^2{\Theta}_{\bf q}  = \bar{A} q^2 \partial_t\delta \rho_{\bf q} -
(iv_1 q_{||} + K q^2) \partial_t \Theta_{\bf q} +
\bar{B}q^2  \Theta_{\bf q}
\eeq
for Fourier components at wavevector ${\bf q}$. It is straightforward to see
that the eigenfrequencies of this problem are of order $q$. Using this fact in
(\ref{rhodiff}) tells us that $q \delta \rho_{\bf q} \sim \Theta_{\bf q}$ to
leading order in $q$. The term $\bar{A} q^2 \partial_t\delta \rho_{\bf q}$ is
thus of the same order as the last term $\bar{B}q^2  \Theta_{\bf q}$. The
presence of an additional conserved field, $\rho$, therefore does not affect
the power-counting properties of the equation for the orientational order
parameter. If we introduce a noise into (\ref{polarfull}) the resulting
fluctuations in the orientation will scale precisely as in the Toner-Tu flock
\cite{tonertu}, and therefore the fluctuations in the concentration of rods,
via (\ref{rodcontord}), will be singular.

A word on the parameter $A$: it appears plausible on packing grounds
that it should be negative, i.e., that the thick end of the rod is easier
accommodated in low-density regions. However, it proved numerically
infeasible to detect density variations on the length scale of a
single rod; we defer further discussion of this coefficient to later work as it
does not play a significant role in our model.

\section{Acknowledgements}SR thanks Narayanan Menon for crucial suggestions
 regarding the role of the bead velocity, Menon and Vijay Narayan for early 
ideas on  bead-rod systems, Rajaram Nityananda for a valuable email
exchange, and Ananyo Maitra and Suropriya Saha for many discussions. For
support, we acknowledge the UGC (NK), the CSIR
(HS), and a DST J C Bose Fellowship (SR and AKS). NK acknowledges the
hospitality of TCIS, TIFR Hyderabad. 

\section{Author contribution} NK and AKS planned the experiments, and NK carried
them out. HS and SR planned the simulations, HS wrote and ran the codes. SR and
HS developed the analytical theory. All authors contributed to interpreting the
results of experiment and simulation and relating them to the theory, and
drafted the paper jointly.

\setcounter{figure}{0}

\makeatletter 
\renewcommand{\fnum@figure}{Supplementary Figure \@arabic\c@figure}

\makeatother

\newpage

\begin{center}
\Large{\textbf{\underline{Supplementary Material}}}
\end{center}
\begin{figure}[H]
\centerline{\includegraphics[width=0.55\textwidth]{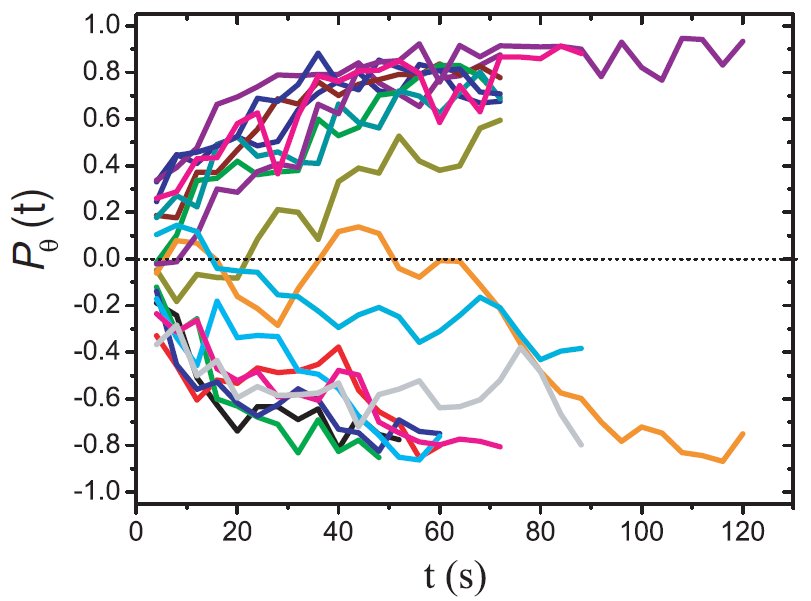}}
\caption{\textbf{Directionality of the flock.} Time evolution of angular component of the rod polar parameter $P_\theta$
for twenty experiments with $\Phi_r=0.05$ and $\Phi_b = 0.68$ showing that
flocks moving clockwise and anti-clockwise occur with equal probability.}
\label{chirality}
\end{figure}

\begin{figure}[H]
\centerline{\includegraphics[width=0.7\textwidth]{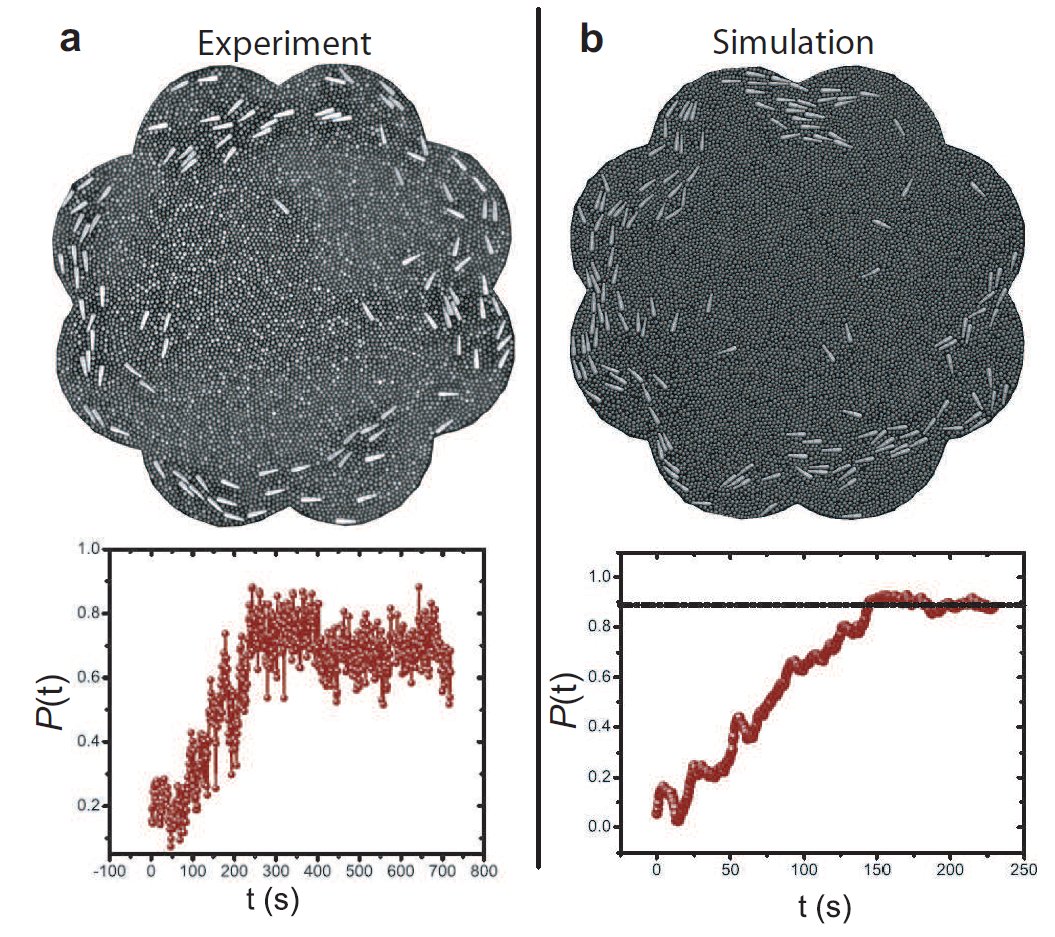}}
\caption{\textbf{Flocking of polar rods does not require a crystalline bead-bed.} (\textbf{a}) An experimental snapshot showing polar particles flocking even in
an amorphous bidisperse medium of bead of diameter 0.8 mm and 1.0
mm, and the corresponding growth curve of $P$. (\textbf{b})
A simulation image and graph of similar behaviour in a polydisperse bead
medium.}
\label{amorphous}
\end{figure}
\begin{figure}[H]
\centerline{\includegraphics[width=0.9\textwidth]{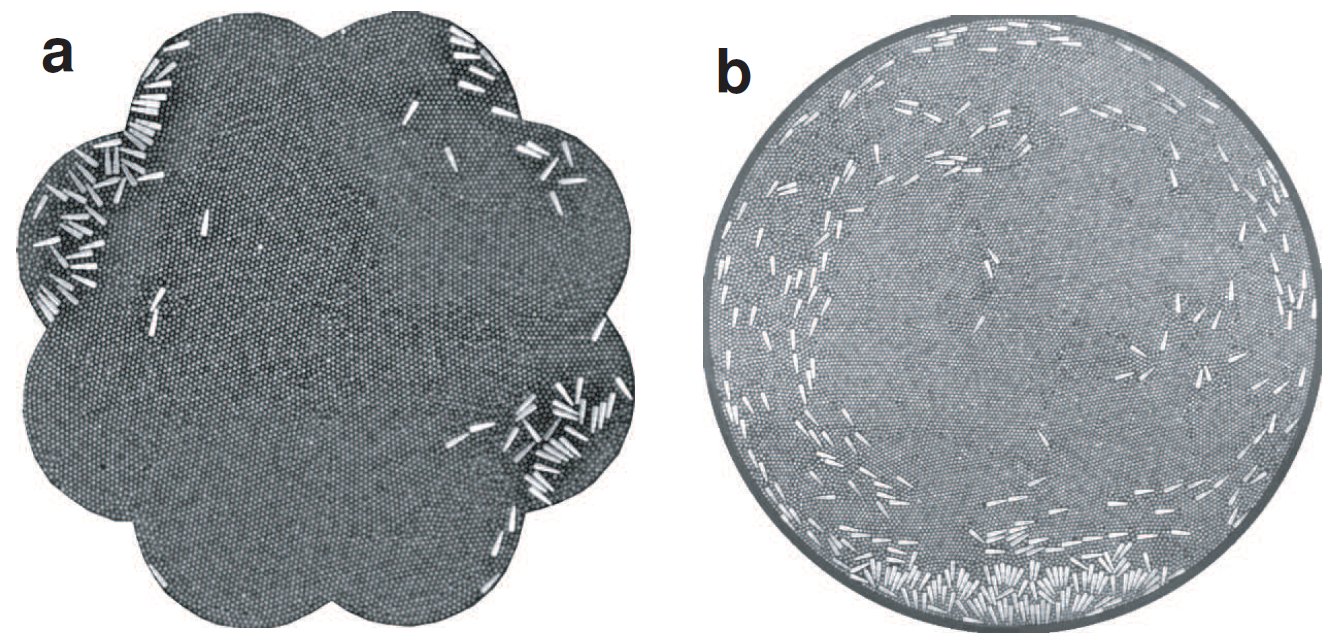}}
\caption{\textbf{Clustering in circular geometry and kinetic arrest in a typical
jammed state.} 
(\textbf{a}) A typical cluster
formation at very high area fractions. Here $\Phi_b=0.70$ and $\Phi_r=0.05$.
(\textbf{b}) Accumulation of the polar rods in circular geometry. Bead and rod
area concentrations are $\Phi_b=0.62$ and $\Phi_r=0.09$ respectively. }
\end{figure}


\begin{figure}[H]
\centerline{\includegraphics[width=1\textwidth]{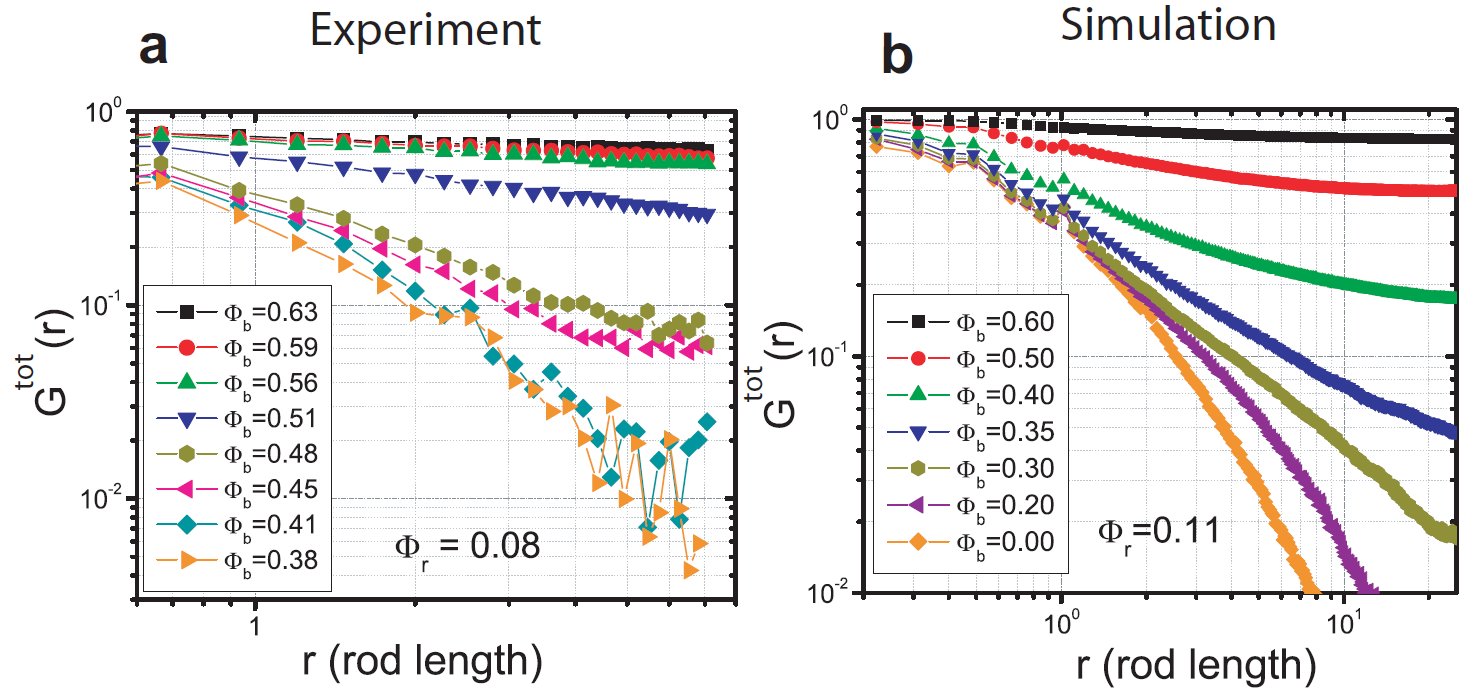}}
\caption{\textbf{Correlation functions of the orientation without the mean subtracted.} $G^{tot}(r) = \langle {\bf p}_i \cdot {\bf p}_j
\rangle_r$ which measures correlations of the orientation \textit{without the
estimated mean subtracted} between pairs $i, j$ of rods with separation $|{\bf
r}_i - {\bf r}_j| = r$. Decay to a nonzero value is clearly seen for high
$\Phi_b$. The smooth evolution as a function of $\Phi_b$ is arguably clearer in
this presentation.}
\label{GrSupp}
\end{figure}

\begin{figure}[H]
\centerline{\includegraphics[width=1\textwidth]{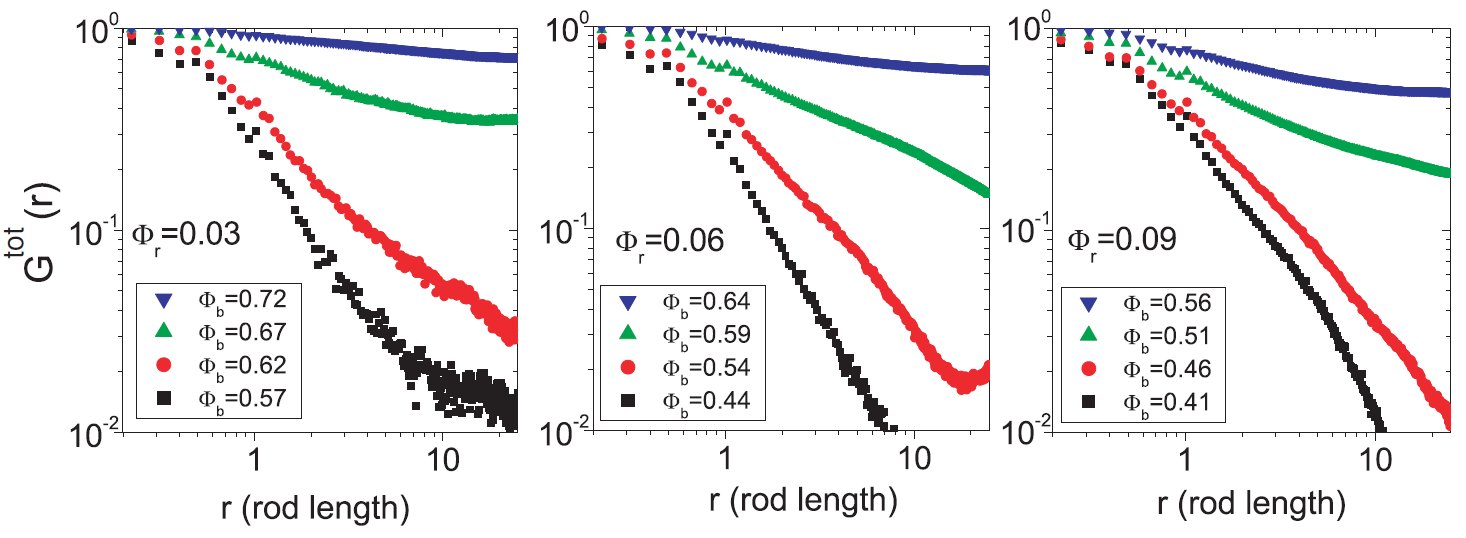}}
\caption{\textbf{Pretransitional fluctuations at very low $\Phi_r$.} Increasing the bead concentration promotes orientational ordering at
rod area fractions as low as $0.03$, as seen in this set of simulation
measurements of the spatial correlations $G^{tot}(r)$ of the polar order
parameter.}
\label{gr_all}
\end{figure}
\begin{figure}[H]
\centerline{ \includegraphics[width=0.9\textwidth]{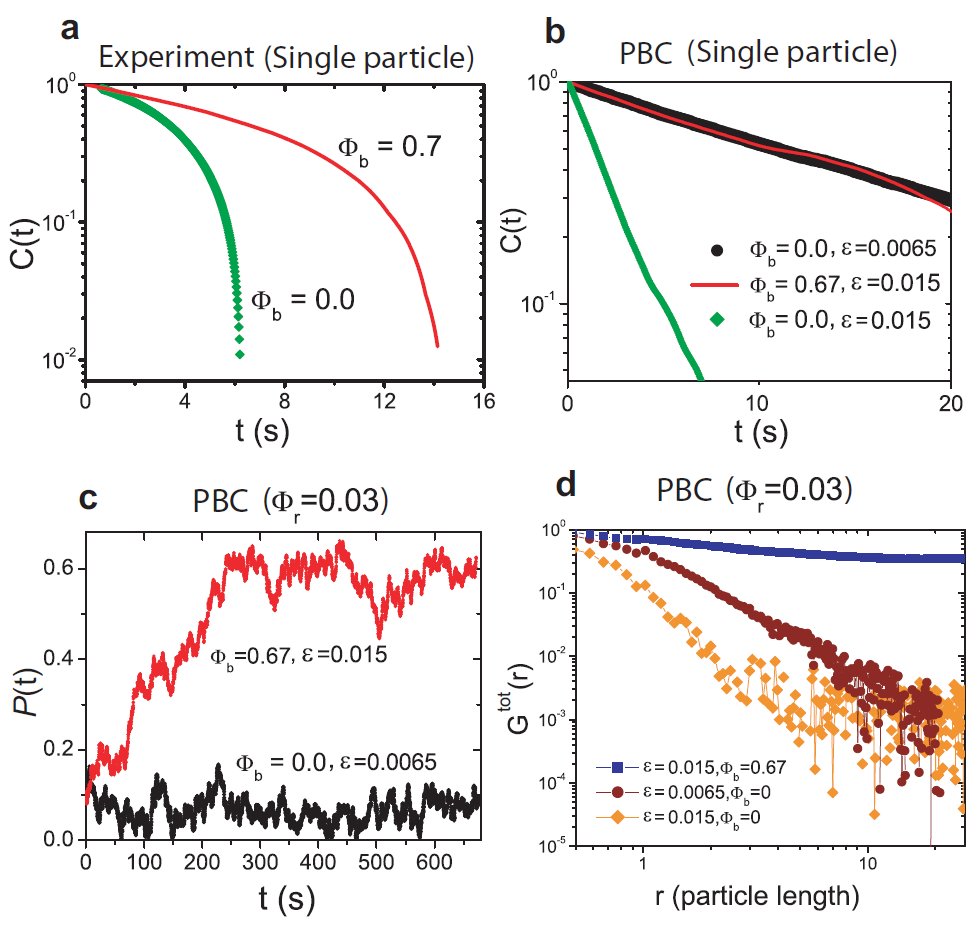}}
\caption{\textbf{Noise suppression by the bead medium.} (\textbf{a}) Comparison of temporal autocorrelation functions of the orientation
of a single polar rod without medium and in a bead medium with $\Phi_b =
0.7$. The medium causes a significant suppression in rotational noise. (\textbf{b}) In
the simulation a single polar particle, with intrinsic rotational noise
strength 0.015 in the absence of a medium, displays an effective rotational
noise strength of 0.0065 when surrounded by a bead medium with $\Phi_b = 0.67$.
(\textbf{c}) Polar rods with noise strength 0.0065 show no ordering at $\Phi_r = 0.03, \,
\Phi_b=0$, although (\textbf{d}) some enhancement of orientational correlations is seen.}
\label{noise}
\end{figure}

\begin{figure}[H]
\centerline{\includegraphics[width=.75\textwidth]{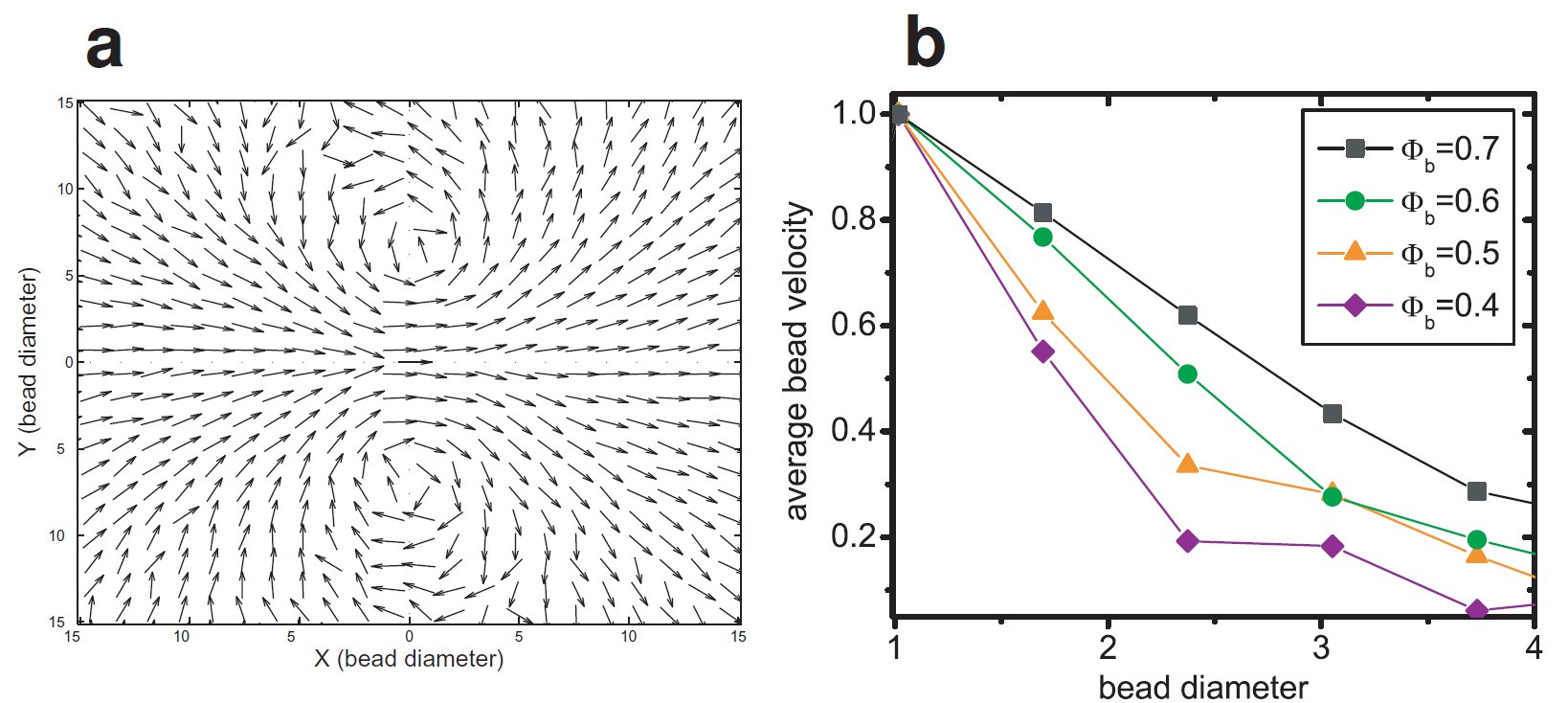}}
\caption{\textbf{Velocity field around a polar rod.} (\textbf{a}) Bead velocity field around a single motile
polar rod in a simulation with $\Phi_b = 0.7$; \textit{unit} vectors are shown
for clarity: the characteristic two-lobe
pattern expected from a monopole force density [1] is clearly seen.
(\textbf{b}) The magnitude of the velocity decays with distance from the rod, but
progressively more slowly at higher $\Phi_b$, consistent with an increasing
viscosity $\eta$ and a constant friction $\Gamma$ with base and lid.}
\label{monopole}
\end{figure}
\section*{Supplementary Notes:}

\subsection*{Supplementary Note
1. Flocking of polar rods does not require a crystalline bead-bed}
Our experiments primarily use monodisperse spherical
beads, inevitably leading to crystallisation of the medium at large $\Phi_b$.
Here, using both experiments and simulation, we show that the flocking can be
observed in an amorphous medium as well. The experiment was conducted in the
medium consisting of beads of sizes 0.8 mm and 1.0 mm at an area fraction
$\Phi_b = 0.66$ divided between the two species, and a constant rod area
fraction $\Phi_r = 0.05$. Supplementary Fig. \ref{amorphous}(a) shows a typical
snapshot of a circulating flock along the boundary and build-up of the order
parameter $P$ with time. A further test is conducted by running a simulation
in a polydisperse bead medium of sizes 0.64, 0.72, 0.8, 0.88 and 0.96 mm at area
fraction $\Phi_b = 0.68$ such that the number of beads is same for all sizes and
$\Phi_r = 0.08$ (see Supplementary Fig. \ref{amorphous}(b)). Clearly an ordered state is seen
along with a rise in $P$ with time implying that a crystalline bead-bed is not
required for flocking.
\subsection*{Supplementary Note
2. Noise suppression by the bead medium}

In order to understand the role played by the medium in the experiment, we
calculate the rotational autocorrelation function $C(t)\equiv \left\langle
\cos(\theta (t+\tau)-\theta (\tau))\right\rangle$ (averaged over time $\tau$)
for
a single polar rod with $\Phi_b = 0$ (i.e., no bead medium) and $\Phi_b = 0.7$.
Supplementary Fig. \ref{noise}(a) shows a significant increase in rotational correlation time
in the presence of the medium suggesting a strong suppression in rotational
noise of the polar rod. For our simulation parameters, with a rotational noise
strength $\epsilon = 0.015$ (see Methods), a single rod in the absence of beads
has a mean rotational relaxation time $\simeq 2.5$ sec, which grows to 13.6 sec
in a bead medium at $\Phi_b = 0.67$, as a result of enhanced rotational friction
due to the beads (see Supplementary Fig. \ref{noise}(b)). This increase in rotational
relaxation time can be achieved without the bead medium by decreasing the noise
strength to 0.0065. In Supplementary Fig. \ref{noise}(c), we compare the behaviour for $\Phi_r
= 0.03, \, \Phi_b=0.67$, with a noise strength of 0.015, with that for $\Phi_r =
0.03, \, \Phi_b=0$, with a noise strength of 0.0065. If the hypothesised noise
suppression were the sole mechanism at work the two should show the same
order-parameter growth kinetics. However, we see a robust growth of order in the
case with beads, while the bead-free low-noise medium shows only a minor
enhancement of spatial correlations (Supplementary Fig. \ref{noise}d). Taken together these
findings establish that the bead medium is
enhancing interactions, not simply lowering noise.

\subsection{Supplementary Note
3. Velocity field around a polar rod}
In our simulation, we study the flow field induced in the bead medium by the
motion of a single polar rod. From Eq. 7 of the main text, we expect that the rod acts
like a localised force density which, in a two-dimensional fluid on a
substrate, should result [1] in a ${\bf v}$ profile like that of a
sedimenting particle at short distances but cut off at scales of order
$(\eta/\Gamma)^{1/2}$. And indeed, Supplementary Fig. \ref{monopole} shows
qualitatively such a
profile for the measured velocity field from a simulation with $\Phi_b = 0.7$.
\section*{Supplementary Reference:}
%
%
[1] Brotto, T. \textit{et al.}, Hydrodynamics of Confined Active Fluids, \textit{Phys Rev Lett} \textbf{110}, 038101 (2013). 


\end{document}